 \documentstyle[axodraw]{cjour}
\begin{document}



\authorrunninghead{Suzuki and Schmidt}
\titlerunninghead{Loop integrals in three outstanding gauges}






\def\be{\begin{equation}}
\def\ee{\end{equation}}
\def\beq{\begin{eqnarray}}
\def\eeq{\end{eqnarray}} 
\def\s{\sigma} 
\def\p{\rho}
\def\pp{{\bf p}}
\def\G{\Gamma} 
\def\F{_2F_1} 
\def\an{analytic} 
\def\ac{\an{} continuation} 
\def\hsr{hypergeometric series representations} 
\def\hf{hypergeometric function} 
\def\ndim{NDIM} 
\def\quarto{\frac{1}{4}} 
\def\half{\frac{1}{2}} 

\title{Loop integrals in three outstanding gauges: \\ Feynman, Light-cone and Coulomb}


\author{Alfredo T. Suzuki, Alexandre G. M. Schmidt}
\affil{Instituto de F\'{\i}sica Te\'orica - UNESP \\ R.Pamplona, 145 S\~ao Paulo - SP CEP 01405-900, Brazil}

\email{suzuki@ift.unesp.br, schmidt@ift.unesp.br}

\abstract{We apply negative dimensional integration method (\ndim ) to
three outstanding gauges: Feynman, light-cone and Coulomb gauges. Our
aim is to show that \ndim{} is a very suitable technique to deal with
loop integrals, being them originated from any gauge choice. In
Feynman gauge we perform scalar two-loop four-point massless
integrals; in the light-cone gauge we calculate scalar two-loop
integrals contributing for two-point functions without any kind of
prescriptions, since \ndim{} can abandon such devices -- this
calculation is the first test of our prescriptionless method beyond
one-loop order; finally, for the Coulomb gauge we consider a four
propagator massless loop integral, in the split dimensional
regularization context.
 } 


\begin{article}

\contents

\section{Introduction}

Perturbative approach for Quantum Field Theory in any gauge deals with Feynman 
diagrams, which are expressed as $D$-dimensional integrals. The success of 
such approach can be understood from the comparison between the $a=\half (g-2)$ 
measure for the electron, 
  \beq && a_{The} = 1159652201.2(2.1)(27.1) 
\times 10^{-12}\nonumber\\
 && a_{Exp} = 1159652188.4(4.3) \times 10^{-12} ,\eeq
see for instance \cite{laporta}.

This is the best motivation for studying Quantum Field Theory, no physical
theory can give such accuracy in any measurement. In other words, it is the
very best we have.

In section 2 we discuss some 2-loop 4-point functions, namely, on-shell double boxes with 5 and 
6 massless propagators; section 3 is devoted to non-covariant gauges: the light-cone and Coulomb ones. The 
integrals we study for the former have 7 propagators (2-loops) and the latter is 1-loop and have 4, however 
it is also complicated since we have to use split dimensional regularization(SDR). In the final section, 4, 
we present our concluding remarks.

\section{Feynman gauge: scalar two-loop four-point massless integrals}

Of course, covariant gauges are the most popular, in what we could call 
``gauge market"\cite{leib-rmp}. Several methods were and are still developed
to evaluate complicated Feynman loop integrals, being them concerned with
analytic or numerical results\cite{laporta,chetyrkin,halliday}, all in the
context of dimensional regularization\cite{bolini}. 

Our work in concerned with the application of negative-dimensional 
integration method (\ndim{}). It is a technique which can be applied to any
gauge, covariant or non-covariant alike. The results are always expressed as
hypergeometric series which have definite regions of convergence allowing one to study the 
referred diagrams or process in specific kinematical regions of external momenta and/or masses.

On the other side, \ndim{} has a drawback: the amazing number of series -- in
the case where one is considering massless diagrams -- which must be summed.
When such sums are of gaussian type, it is quite easy to write a small
computer program that can do the job algebraically. However, when the series
are of superior order, $_{p+1} F_p$, for $p\geq 2$, there are no known
formulas which can reduce it to a product of gamma functions for any value of
its parameters. Despite this technical problem, \ndim{} proved to be an
excellent method\cite{probing,box,without,coulomb}.

A question which is often arised is: what is more difficult to handle, graphs
with more loops or graphs with more legs? In our point of view, i.e., in the
context of \ndim , the greater the number of loops the heftier the
calculations will be needed to solve it. We will consider in this section a diagram which has 
both (great number of legs and loops, four and two respectively),  a scalar two-loop double-box 
integral where all the particles are massless and the external legs are on-shell.

\subsection{Double box with 5 and 6 propagators}
Let us consider the diagram of figure 1. Consider as the generating functional for our 
negative-dimensional integral the gaussian one, where all external legs are on-shell,
\beq G_b &=& \int d^D\!\! q\; d^D\! r\;\;\exp{\left[-\alpha q^2-\beta (q-p)^2 -\gamma(q-p-p')^2 
-\theta (q-r-p_1)^2 - \phi r^2 \right. } \nonumber \\
&& \left. -\omega (q-r)^2 \right], \\ 
&=& \left(\frac{\pi^2}{\Lambda}\right)^{D/2} \exp{\left[ \frac{1}{\Lambda} \left( -\gamma\phi\omega s - 
\beta\theta\phi t\right)\right]}, \eeq
where $(s,t)$ are the usual Mandelstam variables and we use
$s+t+u=0$. Observe that in the particular case where
$\alpha=0$ we recover the gaussian integral for the diagram of figure
2. We also define $\Lambda = \alpha\theta + \alpha\phi + \alpha\omega
+\beta\theta+\beta\phi+ \beta\omega +\gamma\theta + \gamma\phi
+\gamma\omega+ \phi\omega +\theta\phi$.

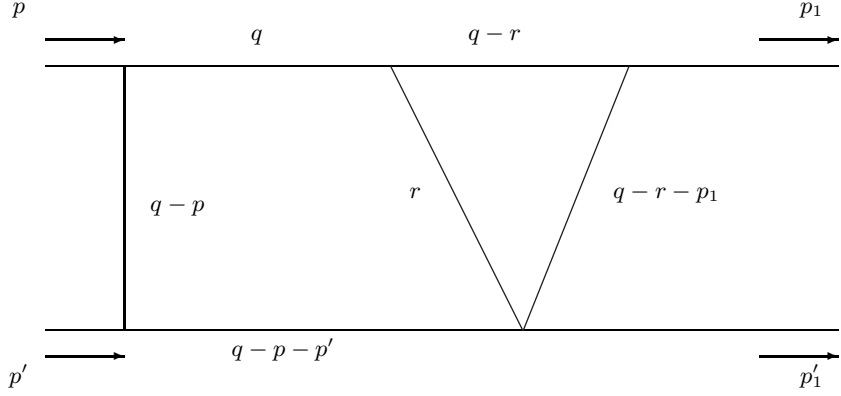
\begin{figure} \label{double-box 6 prop}
\begin{center}
\begin{picture}(600,200)(0,150)
\vspace{40mm}
\thinlines
\put(20,200){\line(1,0){300}}  
\put(20,190){\vector(1,0){30}}\put(20,300){\line(1,0){300}} 
\put(20,310){\vector(1,0){30}}
\put(50,200){\line(0,1){100}} 
\Line(200,199)(150,299)
\Line(200,199)(240,299)
\put(290,190){\vector(1,0){30}}
\put(290,310){\vector(1,0){30}}

{\small 
\put(110,190){\makebox(0,0)[b]{$q-p-p'$}}
\put(310,320){\makebox(0,0)[b]{$p_1$}}
\put(310,180){\makebox(0,0)[b]{$p_1'$}}
\put(10,320){\makebox(0,0)[b]{$p$}} 
\put(10,180){\makebox(0,0)[b]{$p'$}}
\put(100,310){\makebox(0,0)[b]{$q$}}
\put(160,250){\makebox(0,0)[b]{$r$}}
\put(70,245){\makebox(0,0)[b]{$q-p$}}
\put(190,310){\makebox(0,0)[b]{$q-r$}} 
\put(255,250){\makebox(0,0)[b]{$q-r-p_1$}}}
\end{picture}\caption{Double-box with six propagators.}
\end{center}\end{figure}

The usual technique reveals that there are thirteen sums and seven
equations. From the combinatorics one can solve such constraints in
1716 different ways. Of course several systems have no solution -- not
even in the homogeneous case -- and from our previous works we know
that some results are $n$-fold degenerated and others are related by
\ac{}. The result for the integral in question

\beq {\cal BOX} \!\! &=& \!\!\int d^D\!\! q \, d^D\! r\;\; (q^2)^i
(q-p)^{2j} (q-p-p')^{2k} (q-r-p_1)^{2l} (r^2)^m (q-r)^{2n}, \\ 
&=& (-\pi)^D i!j!k!l!m!n!\Gamma(1-\s_b-D/2) \sum_{{\rm all}=0}^\infty
\frac{s^{X_1}t^{X_2}}{X_1! X_2!Y_1!...Y_9!Z_1!Z_2!} {\bf \delta} ,\eeq
where ``all'' means $\{X_1,X_2,Y_1,Y_2,Y_3, Y_4,Y_5,Y_6,Y_7,Y_8, Y_9,Z_1,Z_2\}$ and 
$$ {\cal BOX} = {\cal BOX}(i,j,k,l,m,n), $$ 
must be understood and ${\bf delta}$ represents the system of constraints. The above expression can be expressed, in principle, as a seven-fold hypergeometric series, there are three
possibilities, 
\be {\cal F}(...|z,z^{-1},1), \qquad\qquad {\cal F}(...|z,1), \qquad\qquad {\rm and }\qquad\quad 
{\cal F}(...|z^{-1},1), \ee
where $z=-s/t$. Some series with unit argument, if they were gaussian
can be summed up.  However, a hypergeometric function is meaningful
only if the series which defines it was convergent. Since the first
possibility cannot be convergent we disregard it.

Among the 624 total solutions of the system of constraints we look for
the simplest solution, namely, the one in which we can sum the great number
of series. It is not difficult to find it using computer facilities,
\be\label{6prop} {\cal BOX}^{AC} (i,j,k,l,m,n) = f_1(i,j,k,l,m,n) \; _3F_2
(\{1\}|z) ,\ee
where the five parameters are quoted in the table, $\s_b=
i+j+k+l+m+n+D$ and
\beq f_1(i,j,k,l,m,n) &=& \pi^D (t^2)^{\s_b} (-j|\s_b) (-l|\s_b) (\s_b+D/2|-2\s_b-D/2) \nonumber\\
&\times & (-m|l+m+n+D/2) (i+j+k+m+D|-m-D/2) \nonumber\\ 
&\times & (l+m+n+D|-l-n-D/2), \eeq
besides this one, we can have Appel's, Lauricella's and even more complicated
hypergeometric functions. Moreover, 
$$ (x|y) \equiv (x)_y = \frac{\G(x+y)}{\G(x)}. $$
If we remember that the final result should be
the sum of linearly independent series\cite{box,oleari}, we can rightfully
ask if one is not missing two other $_3F_2$ functions. According to
Luke and Slater\cite{luke}, the differential equation for $_pF_q$ has $p$ linearly
independent solutions, so we should write a sum of three terms. On the other
hand, according to No$\!\!\! /$rlund\cite{norlund}, if the difference between
an upper parameter and a lower one was an integer number, then some series do
no exist --- we used this theorem in \cite{box}.  So, eq.(\ref{6prop}) is the final result for the 
referred integral in the region where $|z|<1$. The expression for the same
graph outside this region can be obtained making the substitutions,

\be s \leftrightarrow t, \qquad j\leftrightarrow  k, \qquad l\leftrightarrow n,
\label{subst} \ee
so we have other $_3F_2$ \hf{} as the result for $|z|>1$.

\begin{table}[h]
\caption{Parameters of hypergeometric functions $\,_3F_2$ representing box integrals}
\begin{tabular*}{\textwidth}{@{\extracolsep{\fill}}lcr}
\hline
Parameters & $\;_3F_2(\{1\}|z) $ & $\;_3F_2(\{2\}|z) $ \cr
\hline
a& $-k$  & $-k$\cr
b& $-n$  & $-n$\cr
c& $-\s_b$ & $-\s_b'$\cr
e& $1+j-\s_b$ & $1+j-\s_b'$\cr
f& $1+l-\s_b$  & $1+l-\s_b'$\cr
\hline
\end{tabular*}
\end{table}

Another solution for the Feynman integral can be written as a  triple hypergeometric series,
\beq {\cal BOX}_3 &=& \pi^D t^j s^{\s_b-j} f_3\sum_{Y_i=0}^\infty\frac{(-j|Y_{456})(m+D/2|Y_{46})(l+n+D/2|Y_5)}{Y_4!Y_5!Y_6!(1+\s_b-j|Y_{456})} \\ &&\times\frac{(i+k+l+m+D|Y_{45}+2Y_6)(i+l+m+n+D|Y_{456})z^{Y_{456}}}{(1-j+l|Y_{56}) (i+j+k+m+D|Y_{46})(l+m+n+D|Y_{456})}, \nonumber\\ 
&& + (j\leftrightarrow l), \nonumber\eeq 
where 
\beq f_3 &=& (-l|j) (l+m+n+D|i) (\s_b+D/2|-j-n-D/2) (-n|l+2n+D/2) \nonumber\\ 
&& \times (i+j+k+m+D|-m-D/2)(-k|j+k-\s_b)(-m|2m+D/2), \eeq 
obviously the above series converges if $|z|<1$, besides other possible
condition on $z$. So there is an overlapping between the regions of convergence of ${\cal BOX}$ and 
${\cal BOX}_3$, so there exists an \ac{} formula which relates both. As fas as we know textbooks do not 
show formulas relating triple hypergeometric series with simple ones.

We have also 4-fold series,
\beq {\cal BOX}_4 &=& \pi^D t^l s^{\s_b-l} f_4 \sum_{Y_i=0}^\infty \frac{(-l|Y_{1247})
(m+D/2|Y_{147})(-i|Y_1)}{Y_1!Y_4!Y_7!Y_2!(1+\s_b-l|Y_{1247})(1+j-l|Y_{127})} \\
&&\times
\frac{(i+l+m+n+D|Y_{47})(i+j+m+n+D|Y_{124}+2Y_7)z^{Y_{1247}} }{
(i+l+m+n+D|Y_{47} )(l+m+n+D|Y_{147})}\nonumber\\
&& + (j\leftrightarrow l), \nonumber \eeq 
where
\beq f_4 &=& (-j|l) (l+m+n+D|i+j-l) (\s_b+D/2|-k-l-D/2)(-m|2m+D/2)\nonumber\\
&&\times (-k|k+l-\s_b) (-n|l+2n+D/2). \eeq 
Observe that the two previous
results are singular when $j-l={\rm integer}$, since we have $\G(j-l)$ or
$\G(l-j)$ in the numerator. However, such singularity cancels if one consider propagators exponents in the analytic regularization context, i.e., introduce\cite{box,davyd} for instance $j=-1+\delta$, then expand the whole expression around $\delta=0$. Proceeding in this way the pole in $\delta$ cancels.  

We have above reduction formulas which
transform a hypergeometric function defined by triple and 4-fold series in a
simpler function defined by a unique sum. These formulas are not in the
textbooks on the subject. It is an original result.

\subsubsection{Double box with 5 propagators} 
The graph of figure 2, is a special case of the previous one. In the gaussian integral 
$\alpha$ must be zero, so in the final result we must merely take $i=0$,

\begin{figure}\label{double-box 5 prop}
\begin{center}
\begin{picture}(600,200)(0,150)
\vspace{15cm}
\thinlines
\put(20,200){\line(1,0){300}}  
\put(20,190){\vector(1,0){30}}
\put(20,300){\line(1,0){300}} 
\put(20,310){\vector(1,0){30}}
\Line(200,199)(150,299)
\Line(200,199)(240,299)
\Line(150,299)(100,199)
\put(290,190){\vector(1,0){30}}
\put(290,310){\vector(1,0){30}}

{\small
\put(140,190){\makebox(0,0)[b]{$q-p-p'$}} 
\put(310,320){\makebox(0,0)[b]{$p_1$}} 
\put(310,180){\makebox(0,0)[b]{$p_1'$}} \put(10,320){\makebox(0,0)[b]{$p$}} 
\put(10,180){\makebox(0,0)[b]{$p'$}} 
\put(160,250){\makebox(0,0)[b]{$r$}}\put(100,245){\makebox(0,0)[b]{$q-p$}} 
\put(190,310){\makebox(0,0)[b]{$q-r$}} 
\put(255,250){\makebox(0,0)[b]{$q-k-r$}}}\end{picture}
\caption{Double-box with five propagators.}
\end{center}\end{figure}
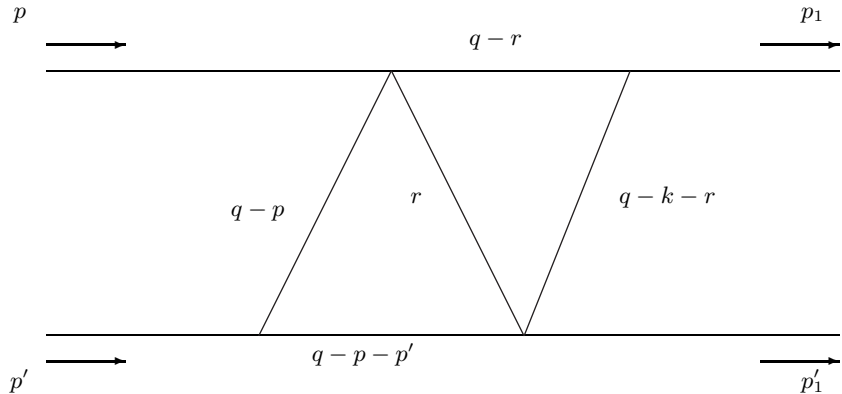

\be\label{5prop} {\cal BOX}^{AC} (0,j,k,l,m,n) = f(0,j,k,l,m,n) \; _3F_2 (\{2\}|z) ,\ee
where the parameters are listed in the table and we define $\s_b'= j+k+l+m+n+D$ and
\beq f(0,j,k,l,m,n) &=& \pi^D (t^2)^{\s_b'} (-j|\s_b') (-l|\s_b') (\s_b'+D/2|-2\s_b'-D/2) \nonumber\\
&&\times (-m|l+m+n+D/2) (j+k+m+D|-m-D/2)\nonumber\\
&&\times (l+m+n+D|-l-n-D/2), \eeq
we can proceed with the same substitutions (\ref{subst}) to obtain the result outside the region $|z|<1$.

Finally, when all the exponents are equal to minus one, the $_3F_2$ colapses
to a $_2F_1$ which can be written as an elementary function.

The results ${\cal BOX}_3$ and ${\cal BOX}_4$, in the same special case
$(i=0)$, are \hsr{} for the integral in question. Since they are different
and depend on the same variable we must sum them in order to get a triple
series representation for ${\cal BOX}_3(0,j,k,l,m,n)$.

\section{Non-covariant gauges: Light-cone and Coulomb}
Recently there have been many works on non-covariant gauges, namely,
light-cone\cite{lc,lc2}, Coulomb\cite{coul} and radial and axial gauges\cite{radial}.
Despite they are not so popular as covariant ones, they have some important
features which can help our study on certain physical problems.

Light-cone gauge, as far as we know, is the only one where certain 
supersymmetric theories can be shown to be UV finite and possess a local
Nicolai map\cite{leib-cjp}. Moreover, ghosts decouple from physical
particles and one is left with a reduced number of diagrams. On the other
hand, the price to pay seemed to be so high, since the gauge boson propagator did 
generate spurious poles in physical amplitudes. This problem was overcame when
 andelstam and Leibbrandt\cite{mandel} introduced causal prescriptions to
treat such poles (there are also other causal prescription which can be
implemented, proposed by Pimentel and Suzuki\cite{pimentel}, known as causal
Cauchy principal value prescription.) However, the famous ML-prescription
necessarily forces one to use partial fractioning tricks and integration over
components, which turn the calculations rather involved\cite{leib-nyeo}. 

Negative-dimensional approach can avoid at all the use of prescriptions and provide physically
acceptable results, i.e., causality preserving results. The calculation we will present is the 
very first test beyond 1-loop order without invoking ML-prescription, as we called in 
\cite{without} \ndim{} is a prescriptionless method. Still, integration
over componenets and partial fractioning tricks can be completely abandoned as
well as parametric integrals. The important point to note\cite{without} is
that the dual light-like 4-vector $n^*_\mu$ is necessary in order to span the needed
4-dimensional space\cite{leib-cjp,tetrad}. This is the very reason why our
calculation for one-degree of covariance violation failed\cite{probing}.

The second non-covariant gauge we deal with in this paper is the Coulomb gauge. Potential
between quarks and studies on confinement are easily performed in this gauge\cite{lc2,coul}. 
Besides, the ghost propagator has
no pole in this gauge! As light-cone gauge, Coulomb also have problems with 
gauge boson propagator. In the former, loop integrals generated aditional
poles; in the latter, such integrals are not even defined\cite{taylor} since
they have the form,
\be \int \frac{dq_4 d^3{\bf q}}{{\bf q}^2} ,\ee
such objects are the so-called energy-integrals. Doust and
Taylor\cite{taylor} presented a solution for this issue in a form of a
interpolating gauge (between Feynman and Coulomb). Leibbrandt and
co-workers\cite{split} presented also a solution, a procedure they called {\it
split dimensional regularization}, (SDR), which introduces two regulating
parameters, one for the energy component and another for the 3-momentum one.
So, the measure becomes,
 \be d^D\! q = dq_4 d^{D-1}\! {\bf q}\;\; \longrightarrow
^{\!\!\!\!\!\!\!\!\!\!\!\! sdr} \;\; d^D\! q = d^{\rho}\! q_4 d^{\omega}\!
{\bf q}, \ee
 in Euclidean space. 

\ndim{} can also deal with Coulomb gauge
loop integrals, but it needs to make use of SDR. In this work we propose to
apply \ndim{} to scalar integrals with four massless propagators. Our results
are given in terms of hypergeometric series involving external momenta,
exponents of propagators and regulating parameters $\omega$ and $\p$.

\subsection{The Light-Cone Gauge}
 So far, we have tested our \ndim{} for
integrals pertaining to one-loop class. Now we apply such technology to some 
massless two-loop integrals. Let us consider an integral studied by Leibbrandt and 
Nyeo\cite{leib-nyeo}, since they did not present the full result for it, 
\be C_3 = \int d^D\! q\, d^D\! k\, \frac{k^2}{q^2 (q-k)^2 (k-p)^2 (k\cdot n) (q\cdot n)}, \ee
where in their calculation ML-prescription must be understood. On the other hand, in the \ndim{} context 
the key point is to introduce the dual vector $n_\mu^*$ in order to span the needed 
space\cite{without,leib-cjp,tetrad}. If we do not consider it, our result will violate 
causality, giving the Cauchy principal value of the integral in question, as we conclude in 
\cite{probing}.

\ndim{} can consider lots of integrals in a single calculation. Our aim to is perform,

\be {\cal N} = \int d^D\! q\, d^D\! k_1\, (k_1^2)^i (q^2)^j (q-k_1)^{2k} (k_1-p)^{2l} 
(k_1\cdot n)^m (q\cdot n)^s (k_1\cdot n^*)^r, \ee 
we will carry out this integral and then present results for special cases, including Leibbrandt and Nyeo's 
$C_3$, where $i=-1$, $r=0$ and the other exponents equal to minus one. Observe that the integral must be 
considered as function of external momentum, exponents of propagators and dimension,
 \be {\cal N} = {\cal N}(i,j,k,l,m,r,s;P,D), \ee
where $P$ represents $(p^2, p^+, p^-, \half (n\cdot n^*))$, and we adopt the usual notation of 
light-cone gauge\cite{leib-rmp}.

Our starting point is the generating function for our negative-dimensional integrals,
\beq G_N &=& \int d^D\! q\, d^D\! k\, \exp{\left[-\alpha k^2 -\beta q^2 -\gamma (q-k)^2 -\theta (k-p)^2 -\phi (k\cdot n) -\omega (q\cdot n) \right. }\nonumber\\
&& \left. -\eta (k\cdot n^*)\right], \eeq
then after a little bit of algebra we integrate it,
\beq G_N &=& \left(\frac{\pi^2}{\lambda}\right)^{D/2} \!\!\!\exp{\left\{ \frac{1}{\lambda}\left[ -g_1p^2 
-g_2(p\cdot n) -g_3 (p\cdot n^*) +g_4 (\half n\cdot n^*)\right] \right\} }, \eeq
where
$$g_1 = (\alpha\beta + \alpha\gamma +\beta\gamma)\theta, \quad g_2 = (\beta\phi+ \gamma\omega +\gamma\phi)\theta,\quad g_3 = (\beta+\gamma)\eta\theta, \quad g_4 = \eta\frac{g_2}{\theta},  $$
and $ \lambda = \alpha\beta+ \alpha\gamma + \beta\gamma + \beta\theta + \gamma\theta.$

Taylor expanding the exponentials one obtain,
\beq {\cal N} &=& (-\pi^D) i!j!k!l!m!r!s!\G(1-\s_n-D/2) \sum_{{\rm all}=0}^\infty \frac{ \delta}{ X_1!\dots X_8!Y_1!Y_2!Y_3!}\nonumber\\
&&\times \frac{(p^2)^{X_{123}}(p^+)^{X_{456}}(p^-)^{X_{78}}}{Z_1!\dots Z_5!}\left( -\frac{n\cdot n^*}{2}\right)^{Y_{123}}, \eeq
where $\s_n=i+j+k+l+m+r+s+D$ and $\delta$ represents the system of constraints $(8\times 16)$ 
for the negative-dimensional integral. In the end of the day we have 12870 possible solutions 
for such system! Most of them, 9142, have no solution while 3728 present solutions which can be written as hypergeometric series. Of course several of these will provide the same series representation, these solutions we call degenerate.

We present a result for the referred integral as a double hypergeometric series,
\beq {\cal N} &=& \pi^D f_n P_n \sum_{Z_j=0}^\infty \frac{(\s_n+D/2|Z_{45}) (i+j+k+m+s+D|Z_{45})(D/2+k|Z_4)}{Z_4!Z_5!(1+i+j+k+\s_n+D|Z_{45})(j+k+s+D|Z_{45})}\nonumber\\
&&\times \frac{(j+s+D/2|Z_5)(i+j+k+r+D|Z_{45})}{(1+i+j+k+m+r+s+D|Z_{45})} \left(\frac{p^2n\cdot n^*}{2p^+p^-}\right)^{Z_{45}}, \eeq
where 
\beq f_n &=& (-m|-s)(-i-j-k-D/2|-\s_n-D/2) (j+k+s+D|i-s+r)\nonumber\\
&&\times\frac{(-l|k+l+D/2) (-k|-j-D/2)(-m|j+m+s+D/2)}{(1+r|-i-j-k-m-r-s-D)}\\
&&\times (-j|-i-k-m-r-s-D),\nonumber \eeq
are the Pochhammer symbols and 
\be P_n = (p^2)^{\s_n+i+j+k+D} (p^+)^{l+m+s-\s_n}(p^-)^{l+r-\s_n} \left(\frac{n\cdot n^*}{2} \right)^{\s_n-l}. \ee

Now we can consider the special case ($i=1, j=k=l=m=s=-1, r=0$), studied in \cite{leib-nyeo}, 
 
\beq\label{caso-especial} {\cal N}_{SC} &=& \pi^D \frac{\G(5-2D) \G(D-1) \G(D/2-1) 
\G(2-D/2)\G(D/2-2)}{\G(1-D/2)\G(D-3)} (p^2)^{2D-5} \nonumber\\
&&\times (p^+)^{1-D} (p^-)^{3-D} \left(\frac{n\cdot n^*}{2} 
\right) ^{D-3} \sum_{Z_4,Z_5=0}^\infty \frac{(3D/2-4|Z_{45})}{Z_4!Z_5!(2D-4|Z_{45})}\nonumber\\
&&\times \frac{(D/2-1|Z_4)(D/2-2|Z_5)(D-1|Z_{45})}{(D-2|Z_{45})} \left(\frac{p^2 n\cdot n^*}{p^+p^-} 
\right)^{Z_{45}} ,\eeq 
observe that it exibits a double pole, as stated by Leibbrandt and Nyeo\cite{leib-nyeo}.  

\subsection{The Coulomb Gauge}
 We will present the full calculation of an integral which has four
propagators,\be J(i,j,k,m) = \int d^D\! q\; (q^2)^i(q-p)^{2j}{\bf q}^{2k}
({\bf q+p})^{2m}, \ee
in order to regulate the possible divergences originated by the energy component, SDR must be 
understood, namely, 
 \be d^D\! q = d^{\rho}\! q_4 d^{\omega}\! {\bf q}, \ee
where $D=\p+\omega$.

The generating functional for our negative-dimensional integrals is the gaussian-like integral,
\be G_c = \int d^D\! q\, \exp{\left[-\alpha q^2-\beta (q+p)^2 -\gamma {\bf q}^2
-\theta ({\bf q+p})^2\right]}, \ee
which can be easily integrated,
\be G_c = \frac{\pi^{D/2}}{\lambda_1^{\p/2}\lambda_2^{\omega/2}} \exp{\left( -\frac{\alpha\beta}{\lambda_1} 
p_4^2\right)} \exp{\left[-\frac{(\alpha+ \gamma)(\beta+ \theta)}{\lambda_2}\pp ^2 \right] }. \ee
There are results given by double, triple, 4-fold and 5-fold hypergeometric series in the variable 
$\pp^2/p_4^2$ or its inverse.

We will present two of such \hsr{}, the first one is a 4-fold series,
\beq \label{4-sum} J_4(i,j,k,m) &=& C_4(i,j,k,m) \sum_{X_i=0}^\infty \frac{ (-i|X_{1234}) 
(j+m+D/2|X_{34})}{X_1!X_2!X_3!X_4!} \left(\frac{\pp^2}{p_4^2}\right)^{X_{1234}}  \nonumber\\
&&\times \frac{(-1)^{X_3} (1+j+m+\p/2|X_3-X_{12} )(-m|X_2)}{(1+j+k+m+D/2|X_{34})}\nonumber\\
&& \times \frac{(-j-\p/2|X_1-X_3)(k+\omega/2|X_{124})}{(1-i-\p/2|X_{124})}\nonumber\\ 
&& + (i\leftrightarrow j, k\leftrightarrow m), \eeq
where 
\beq C_4(i,j,k,m) &=& \pi^{D/2} (p_4^2)^i (\pp^2)^{\s_c-i} (-j|-\p/2) (-j-m-\p/2|-k-\omega/2) \nonumber\\
&&\times (-k|2k+\omega/2)(j+k+m+D/2+\omega/2|-k-\omega/2), \eeq
where $\s_c=i+j+k+m+D/2$ and the second a 5-fold hypergeometric series,
\beq \label{5-sum} J_5(i,j,k,m) &=& C_5(i,j,k,m) \sum_{X_i=0}^\infty 
\frac{ (-i-j-\p/2|2X_1+ X_{2345})}{X_1!X_2!X_3!X_4!X_5!} \nonumber\\
&&\times \left(\frac{\pp^2}{p_4^2}\right)^{2X_1+X_{2345}} \frac{(-1)^{X_5} 
(m+\omega/2|X_{1345})}{(1+k+m+\omega/2|X_{145})}\nonumber\\
&& \times \frac{(k+\omega/2|X_{1245})}{(1-j-\p/2|X_{135})(1-i-\p/2|X_{124})}, \eeq
where
\beq C_5(i,j,k,m) &=& \pi^{D/2}(p_4^2)^{i+j+\p/2} (\pp^2)^{k+m+\omega/2} (-i|i+j+\p/2) (-j|i+j+\p/2) 
\nonumber\\
&&\times (-k|k+m+\omega/2)(-m|k+m+\omega/2) (i+j+\p|-\s_c-\p/2)\nonumber\\
&& \times (k+m+\omega|-\s_c-\omega/2), \eeq
observe that the above result is also symmetric in $(i\leftrightarrow j, k\leftrightarrow m)$, which means 
in the loop integral, $ q^\mu \rightarrow q^\mu+p^\mu$.



Another important point to observe is that the final result must a sum of linearly independent 
hypergeometric series\cite{probing,box}. The above 5-fold series, $J_5$, appears only one time 
whereas $J_4$ is degenerate since several systems give its two hypergeometric functions. This 
must be considered if one wants to apply \ndim{} to more complicated diagrams which can in 
principle generate \hsr{} even more involved.

Moreover, the above expressions, $J_4$ and $J_5$, are related by direct \ac , since both are 
convergent for $|\pp^2/p_4^2|<1$. When one is considering simple \hf{}, several formulas are 
known; on the other hand, for rather complicated hypergeometric series, as we obtained --- four 
and five-fold series ---, there are very few of such formulas. \ndim{} can fill this gap, since 
it is the only method which provides \hsr{} for Feynman loop integrals, in different kinematical 
 regions, and related by \ac{} direct or indirect alike.

The above hypergeometric series (only the series, not the factors!), $J_4$ and $J_5$, can be written as 
generalized \hf{}s \cite{luke} of four and five variables,
\beq & & {\cal F}\, ^{6:0;0;0;0}_{2:0;0;0;0} \left[ \matrix{ (-i:1,1,1,1), 
(j+m+D/2:0,0,1,1) (1+j+m+\p/2:-1,-1,1,0)\cr (1+j+k+m+D/2:0,0,1,1) }\right. \nonumber\\
&& \left.\left.\matrix{(-m:0,1,0,0) \cr
(1-i-\p/2:1,1,0,1)}\right|x,x,-x,x,\right], \eeq
and 
\beq & & {\cal F}\, ^{3:0;0;0;0;0}_{3:0;0;0;0;0} \left[ \matrix{ 
(-i-j-\p/2:2,1,1,1,1),(m+\omega/2:1,0,1,1,1)\cr (1+k+m+\omega/2:1,0,0,1,1), (1-j-\p/2:1,0,1,0,1)  } 
\right. \nonumber\\
&& \left.\left.\matrix {(k+\omega/2:1,1,0,1,1)\cr (1-i-\p/2:1,1,0,1,0)} \right| 
x^2,x,x,x,-x\right] ,\eeq
where $x=\pp^2/p_4^2$.


\section{Conclusion}
The technique of Feynman parametrization
can of course be used to perform loop integrals in different gauges but it is
very difficult to perform the parametric integrals for arbitrary exponents of
propagators. Not so with \ndim{}, carry loop integrals out with particular
exponents is as easy as dealing with arbitrary ones -- besides, one can come
across with singularities which depend on them and not on dimension $D$ --
this fact is very important when we are studying light-cone gauge Feynman
integrals, because one could have to handle products like $(q^+)^a
\left[(q-p)^+\right]^b$, being $a$ and $b$ negative. \ndim{} can calculate all
of them simultaneously, but if one chooses partial fractioning tricks then
he/she will be forced to carry out each integral separately. Besides usual
covariant integrals and the trickier light-cone gauge ones, \ndim{} was probed
in the Coulomb gauge, where a procedure -- introduced by Leibbrandt and
co-workers -- called {\it split dimensional regularization} is needed in order
to render the energy integrals well-defined. 

In this paper, we studied
Feynman loop integrals pertaining to three outstanding gauges: the usual, and 
more popular, covariant Feynman gauge and two of the trickiest non-covariant
gauges, the light-cone and the Coulomb ones. Our results are given in terms of
hypergeometric functions and in the dimensional regularization context.

\begin{acknowledgment} AGMS gratefully acknowledges FAPESP (Funda\c c\~ao de 
Amparo \`a Pesquisa de S\~ao Paulo) for financial support.
\end{acknowledgment}

\end{article}
\end{document}